\documentclass[prb,twocolumn,showpacs,preprintnumbers,amsmath,amssymb]{revtex4}
\usepackage{graphicx}
\usepackage{dcolumn}
\usepackage{bm}

\begin{document}

\title{Role of the conduction electrons in mediating exchange
interactions in Heusler alloys}

\author{E. \c Sa\c s\i o\u{g}lu$^{1,2}$}\email{e.sasioglu@fz-juelich.de}
\author{L. M.  Sandratskii$^{1}$}\email{lsandr@mpi-halle.de}
\author{P. Bruno$^{1}$}\email{bruno@mpi-halle.de}

\affiliation{$^{1}$Max-Planck-Institut f\"ur Mikrostrukturphysik,
Weinberg 2, D-06120 Halle, Germany\\
$^{2}$Institut f\"{u}r Festk\"{o}rperforschung, Forschungszentrum
J\"{u}lich, D-52425 J\"{u}lich, Germany}

\date{\today}

\begin{abstract}
Because of large spatial separation of the Mn atoms in Heusler
alloys ($d_{Mn-Mn}>4$ \AA) the Mn 3\textit{d} states belonging to
different atoms do not overlap considerably. Therefore an indirect
exchange interaction between Mn atoms should play a crucial role
in the ferromagnetism of the systems. To study the nature of the
ferromagnetism of various Mn-based semi- and full-Heusler alloys
we perform a systematic first-principles calculation of the
exchange interactions in these materials. The calculation of the
exchange parameters is based on the frozen-magnon approach. The
Curie temperature is estimated within the mean-field
approximation. The calculations show that the magnetism of the
Mn-based Heusler alloys depends strongly on the number of
conduction  \textit{sp} electrons, their spin polarization and the
position of the unoccupied Mn 3\textit{d} states with respect to
the Fermi level. Various magnetic phases are obtained depending on
the combination of these characteristics. The magnetic phase
diagram is determined at zero temperature. The results of the
calculations are in good agreement with available experimental
data. The Anderson's \textit{s-d} model is used to perform a
qualitative analysis of the obtained results. The conditions
leading to diverse magnetic behavior are identified. If the spin
polarization of the conduction electrons at the Fermi energy is
large and  the unoccupied Mn 3\textit{d} states lie well above the
Fermi level, an RKKY-type ferromagnetic interaction is dominating.
On the other hand, the contribution of the antiferromagnetic
superexchange becomes important if unoccupied Mn 3\textit{d}
states lie close to the Fermi energy. The resulting magnetic
behavior depends on the competition of these two exchange
mechanisms. The calculational results are in good correlation with
the conclusions made on the basis of the Anderson \textit{s-d}
model which provides useful framework for the analysis of the
results of first-principles calculations and helps to formulate
the conditions for high Curie temperature.

\end{abstract}


\maketitle

\section{introduction}

In recent years,  Heusler  alloys have become the subject of
intensive experimental and  theoretical investigations.
The strong interest to these systems is mainly due to two unique
properties: half-metallic behavior and
martensitic phase transformations. \cite{Heusler_1,Heusler_2}  The
half-metallicity was first predicted by de Groot and collaborators
in 1983 when studying the band structure of a half-Heusler alloy
NiMnSb.\cite{H_1} Then the half-metallic ferromagnets have become
one of the most studied classes of
materials.\cite{H_2,H_3,H_4,H_5,H_6,H_7,H_8,H_9,H_10,H_11,H_12,H_13}
The existence of a gap in the minority-spin band structure leads
to the 100\% spin polarization of the electron states at the Fermi
level and makes these systems attractive for applications in the
emerging field of spintronics. Besides strong spin polarization of
the charge carriers, the half-metallic
materials should have a crystal structure compatible with the
industrially used zincblende semiconductors and possess a high
Curie temperature to allow the applications in the devices
operating at room temperature. The available experimental information
shows that Heusler alloys are promising materials also in this
respect. \cite{Exp_1,Exp_2,Exp_3,Exp_4}

At low temperatures several Heusler compounds (i.e., Ni$_2$MnGa,
Co$_2$NbSn) undergo a structural  transformation from a highly
symmetric cubic austenitic phase to a low symmetry martensitic
phase. The compounds that are magnetic in the martensitic phase
can exhibit two unusual effects: magnetic shape memory (MSM)
effect and inverse magneto-caloric effect.\cite{MSM_1,IMCE_1} In
MSM alloys an external magnetic field can induce large strains
when applied in the martensitic state. The MSM alloys are  of
great interest as promising smart materials for future
technological applications.\cite{MSM_1,MSM_3,MSM_4,MSM_5} They can
be used as sensors and actuators in different fields of
applications. The inverse magnetocaloric effect (IMCE) has its
origin in martensitic phase transformation that modifies the
exchange interactions due to the change in the lattice
characteristics. The inverse MCE effect was reported for samples
with compositions close to Ni$_2$MnZ ($\textrm{Z}=\textrm{Ga}$,
Sn).\cite{IMCE_1,IMCE_2, IMCE_3,IMCE_4,IMCE_5} In the martensitic
phase, an adiabatic application of a magnetic field, rather than
removal of the field as  in ordinary MCE, causes the sample to
cool. This feature  is regarded promising for the development of
economical and ecological refrigerants working near room
temperature as an alternative to conventional vapor-cycle
refrigeration.

Besides being promising materials for various applications the
Heusler alloys constitute a class of systems that is important
also for fundamental researches. A wide diversity of magnetic
properties makes Heusler alloys critical test systems for the
theoretical models of exchange interactions. Indeed, within the
same family of alloys one finds very different magnetic behavior:
itinerant and localized magnetism, ferrimagnetism,
antiferromagnetism, helimagnetism and other types of non-collinear
ordering.\cite{Heusler_3,Heusler_4,Heusler_5,Heusler_6,Heusler_7,Heusler_8,
Heusler_9}

Despite a key role of the magnetism
in the properties of Heusler alloys
experimental and theoretical studies of the exchange interactions
in Heusler alloys are still rare. The first important information
on the exchange coupling in these systems was provided by the
inelastic neutron scattering experiments of Noda and Ishikawa, and
Tajima \textit{et. al.}, in the late seventies
\cite{Heusler_Noda,Heusler_Tajima}. The authors measured the spin
wave spectra of Ni$_2$MnSn, Pd$_2$MnSn and Cu$_2$MnAl for various
directions in the Brillouin zone and analyzed the results of the
measurements within the  Heisenberg model. They obtained a
long-ranged and oscillatory behavior of the exchange interactions.
The oscillations were reaching beyond the eight nearest neighbors
in all three compounds. This behavior of the exchange interactions
was considered as an evidence for an indirect exchange coupling,
mediated via conduction electrons. The results were interpreted
using either an RKKY model or \textit{s-d} mixing model of
Anderson (double resonance
model).\cite{RKKY,Anderson_1,Anderson_2,Caroli} The double
resonance model was found to be more relevant for Heusler alloys
due to strong mixing of the Mn 3\textit{d}  states with the
conduction electron states of non magnetic 3\textit{d} and
\textit{sp} atoms.

This initial attempt of the theoretical analysis did not succeed
in the description of the sign and the magnitude of exchange
interactions between nearest and  next nearest neighbors in
contrast to good agreement for larger
distances.\cite{Heusler_Malmstrom_1} The failure of the theories
to describe near-neighbor exchange interactions was attributed to
the asymptotic approximations. Price has shown that Anderson
\emph{s-d} mixing model free of asymptotic approximations was able
to capture qualitative features of the observed spin wave spectra
of Pd$_2$MnSn. \cite{Heusler_Price} Malmstr\"om  and Geldart
discussed the effect of  the finite spin distribution around Mn
atoms on the RKKY interactions. \cite{Heusler_Malmstrom_2} The
authors showed that, in spite of simplified treatment of electron
band structure, finite spin distribution could bring the
calculated values  of the exchange interactions in agreement with
the exchange interactions determined for the Ni$_2$MnSn and
Pd$_2$MnSn compounds from the experiments.

An important feature of the model-Hamiltonian approaches is the
possibility of separate study of different exchange mechanisms.
However, the use of adjustable parameters and strong
simplification of the band-structure strongly restrict the
possibility of reliable predictions for concrete materials. The
development of the parameter-free density functional theory (DFT)
has played a crucial role in the understanding of the physical
properties of itinerant-electron ferromagnets.

The first contribution to the density functional theory of the
exchange interactions and Curie temperature in Heusler alloys was
made in an early paper by K\"ubler and collaborators where the
microscopic mechanisms of the magnetism of these systems were
discussed on the basis of the comparison of the ferromagnetic and
antiferromagnetic configurations of the Mn
moments.\cite{Heusler_Kubler} By analyzing the Mn 3\textit{d}
density of states for different magnetic configurations the
authors proposed the mechanism of an indirect exchange coupling
between Mn moments. Recently, the studies of the interatomic
exchange interactions and Curie temperatures in Heusler compounds
were reported by the present authors and Kurtulus \textit{et al.}
\cite{ES_1a,ES_1b,ES_2,ES_3,Kurtulus}

\section{Aims and structure of the paper}

Our previous studies on experimentally well established Ni-based
compounds Ni$_2$MnZ ($\textrm{Z}= \textrm{Ga}$, In, Sn, Sb)
revealed a complex character of the magnetism in these
systems.\cite{ES_1a} In particular, the obtained long range and
oscillatory behavior of the exchange interactions as well as their
strong dependence on the \textit{sp} atom (Z) gave an evidence for
the conduction-electron mediated exchange interactions in Heusler
alloys. The importance of the \textit{sp}-electrons in the
formation of the magnetic properties has been demonstrated by
first-principles calculations also for a number of other
systems.\cite{MnAs,FeRh,Canted} The experimental studies as well
pointed out to the important role of the  \textit{sp}-electrons.
The early measurements by Webster \textit{et al.} have shown this
for quaternary Heusler alloys Pd$_2$MnIn$_{1-x}$Sn$_{x}$ and
Pd$_2$MnSn$_{1-x}$Sb$_{x}$.\cite{Motiv_1} Recent experimental
studies on the Mn-based semi Heusler compounds
Ni$_{1-x}$Cu$_{x}$MnSb and AuMnSn$_{1-x}$Sb$_{x}$ revealed
similarities in the behaviour of the two systems with the
variation of the number of valence
electrons.\cite{Motiv_2,Motiv_3} In particular, the Curie
temperatures of both systems decrease by the same amount with
increasing concentration $x$ and both systems have similar $T_C$
values for compositions with equal numbers of valance electrons.
These studies have shown that both the non-magnetic
3\textit{d}-atoms (X) and the \textit{sp}-atoms (Z) play important
role in establishing magnetic properties. The explanation of the
observed trends in magnetic characteristics of semi-Heusler alloys
requires consideration of competing exchange mechanisms
contributing to the formation of the magnetic state. One of the
important aims of this work is to show that the results of the
parameter-free DFT calculations for many systems can be
qualitatively interpreted in terms of the competition of two
exchange mechanisms.

In contrast to our previous studies, here we go beyond the
stoichiometric compositions and treat several Mn-based semi and
full Heusler alloys within virtual crystal approximation in order
to understand the dependence of physical characteristics on the
valance electron number.

The qualitative interpretation of the calculational results is
based on the analysis of the Anderson's \textit{s-d} mixing
model.\cite{Anderson_1} Since the atomic Mn moments are well
defined the application of the model is well founded. For the last
four decades the Anderson model has been successfully applied to
variety of problems of condensed matter physics and contributed to
better understanding of numerous experimental properties. The
systems studied within the Anderson model vary from dilute 3\textit{d}
magnetic impurities in non-magnetic
metals\cite{Anderson_3,Anderson_4,Anderson_5,Anderson_6} to
complex systems like Heusler
alloys\cite{Heusler_Malmstrom_1,Heusler_Price,Heusler_Malmstrom_2},
rare earths\cite{RE_Falicov,RE_Monachesi}, magnetic
multilayers\cite{ILEC_1,ILEC_2,ILEC_3} and diluted magnetic
semiconductors.\cite{MSC_1,MSC_2,MSC_3,MSC_4}

In the present paper we show that the diversity of magnetic
behavior in Mn-based Heusler alloys can be interpreted in terms of
the competition between ferromagnetic RKKY-type exchange and
antiferromagnetic superexchange. A special attention is devoted to
the role of the \textit{sp} and non-magnetic 3\textit{d} atoms  in
mediating exchange interactions between Mn atoms. The spin
polarization of the conduction electrons appears to be one of the
key parameters in the formation of magnetic characteristics. We
obtain strong correlation between spin polarization of the
\textit{sp} electrons and the strength of the exchange
interactions and, as a result, the value of the Curie temperature.
%The discussion of the influence of the unoccupied states lying
%close to the Fermi level on the strength of the superexchange
%interaction is given. 
It is shown that  the position of unoccupied
peaks of the Mn 3\textit{d} density of states plays an important
role in the determining the value of the antiferromagnetic
superexchange coupling while the
properties of the conduction-electron states at the Fermi level
are mainly responsible for the ferromagnetic RKKY-type exchange
interaction. We compare the influence of non-magnetic 3\textit{d}
versus 4\textit{d} atoms on the exchange coupling  by considering
the stoichiometric Ni$_2$MnZ and Pd$_2$MnZ (Z$=$In, Sn, Sb, Te).

One of the main conclusions of the present treatment is that
numerous features of the magnetism of the Mn-based semi and full
Heusler alloys with different chemical composition can be
described with few parameters. Therefore, the tuning of these
parameters can be considered as a tool in the fabrication of the
materials with desired magnetic properties. 

A part of the present
work has been published elsewhere.\cite{ES_3} The remaining of the
paper is organized as follows. In Section III we present the
details of the calculational approach. Section IV contains the
discussion of the density of states and magnetic moments. In
section V we dwell on the exchange coupling mechanisms and Section
VI gives the conclusions.

\begin{figure}
\includegraphics[scale=0.28]{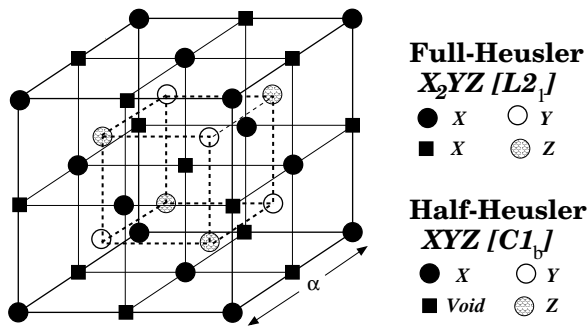}
\caption{$C1_b$ and $L2_1$ structures adapted by the half- and
full-Heusler alloys. The lattice consists from 4 interpenetrating
fcc lattices. In the case of the half-Heusler alloys (XYZ) one of
the four sublattices is vacant. In VCA, the Z site is occupied by
a pseudoatom with a fractional number of valence electrons.}
\label{fig1}
\end{figure}

\section{Technical details}

Semi- and full-Heusler alloys crystallize in the $C1_b$ and $L2_1$
structures respectively (Fig. \ref{fig1}). The lattice
consists from 4 interpenetrating fcc lattices. In the case of the
semi-Heusler alloys (XYZ) one of the four sublattices is vacant.
The Bravais lattice is in both cases fcc.

The calculations are carried out with the augmented spherical wave
method (ASW)\cite{asw} within the atomic-sphere approximation
(ASA).\cite{asa} The exchange-correlation potential is chosen in
the generalized gradient approximation. \cite{gga} A dense
Brillouin zone (BZ) sampling $30\times30\times30$ is used. The
radii of all atomic spheres are chosen equal. In the case of
semi-Heusler alloys we introduce an empty sphere located at the
unoccupied site.

We focuss on the systems where the total magnetic moment is
confined to Mn sublattice that simplifies the interpretation of
the obtained results. To this end we consider Pd  and Cu
containing Mn-based  semi and  full Heusler alloys that can be
written in a compact form as follows
\begin{equation}\nonumber
\textrm{X}_{\textrm{k}}\textrm{MnZ}_{1-\textrm{m}}\textrm{Z}^{\prime}_{\textrm{m}}
\Rightarrow \bigg\{\begin{array}{c}
0\leq \textmd{m} \leq 1, \quad \textrm{k}=1 \quad \textrm{for s-H} \\
0\leq \textmd{m} \leq 1, \quad \textrm{k}=2 \quad \textrm{for f-H}
\end{array}\bigg\} \\ \nonumber
\end{equation}
where  $\textrm{X}=(\textrm{Pd}, \textrm{Cu})$,
$(\textrm{Z},\textrm{Z}^{\prime})=(\textrm{In}, \textrm{Sn})$,
(Sn, Sb), (Sb, Te) and s-H (f-H) stands for semi-Heusler
(full-Heusler) alloys. To account for non-integer electron numbers
we use the virtual crystal approximation (VCA).\cite{VCA_1} In
VCA, the Z site occupied by In and Sn atoms according to their
concentration which is described by an atom with fractional number
of electrons $(1-m)\, z^{In}+ m \, z^{Sn}$ where $z^{Sn}$ is the
number of electrons of Sn and similarly for other chemical
elements. In this approximation only the spin magnetic moment and
the density of states (DOS) of the pseudoatom can be calculated.
The properties of the pseudoatom cannot be projected on the
constituting atoms. An advantage of VCA is the possibility of a
continuous variation of the electron number without resorting to
large super cells.

\begin{table}
\caption{Experimental and theoretical lattice parameters in
X$_{\textrm{k}}$MnZ ( $\textrm{X}=\textrm{Pd}$, Cu; $\textrm{k}=1,
2$; $\textrm{Z}=\textrm{In}$, Sn, Sb, Te). For explanations see
text.}
\begin{ruledtabular}
\begin{tabular}{lccccc}
Compound  & a$_{[\textrm{In}]}$(\AA) & a$_{[\textrm{Sn}]}$(\AA) &
a$_{[\textrm{Sb}]}$(\AA) & a$_{[\textrm{Te}]}$(\AA) &
a$_{[\textrm{Z}]}$(\AA)  \\ \hline
           PdMnZ             &      -        &    -          &      6.25     &  6.27    &      6.26   \\
           CuMnZ             &      -        &    -          &      6.09     &   -      &      6.09   \\
           Pd$_2$MnZ         &     6.37      &   6.38        &      6.41     &   -      &      6.38   \\
           Cu$_2$MnZ         &     6.20      &   6.17        & -             &   -      &      6.18   \\
\end{tabular}
\end{ruledtabular}
\label{table1}
\end{table}

In the  last  column  of Table~\ref{table1}  we  present the
lattice parameters used in the calculations. The remaining columns
give the lattice constants of the experimentally existing systems.
Some of the compounds are not yet synthesized. For the synthesized
systems, the alloys containing the \textit{sp}-atom (Z) from the
same row of the Periodic Table have similar values of lattice
constants. Therefore, we use the average lattice parameters of the
synthesized systems in the calculations for all Z constituents. As
seen from Table~\ref{table1} the difference between experimental
lattice parameters and the parameters used in the calculations is
less than 0.5\%.

\begin{figure*}[t]
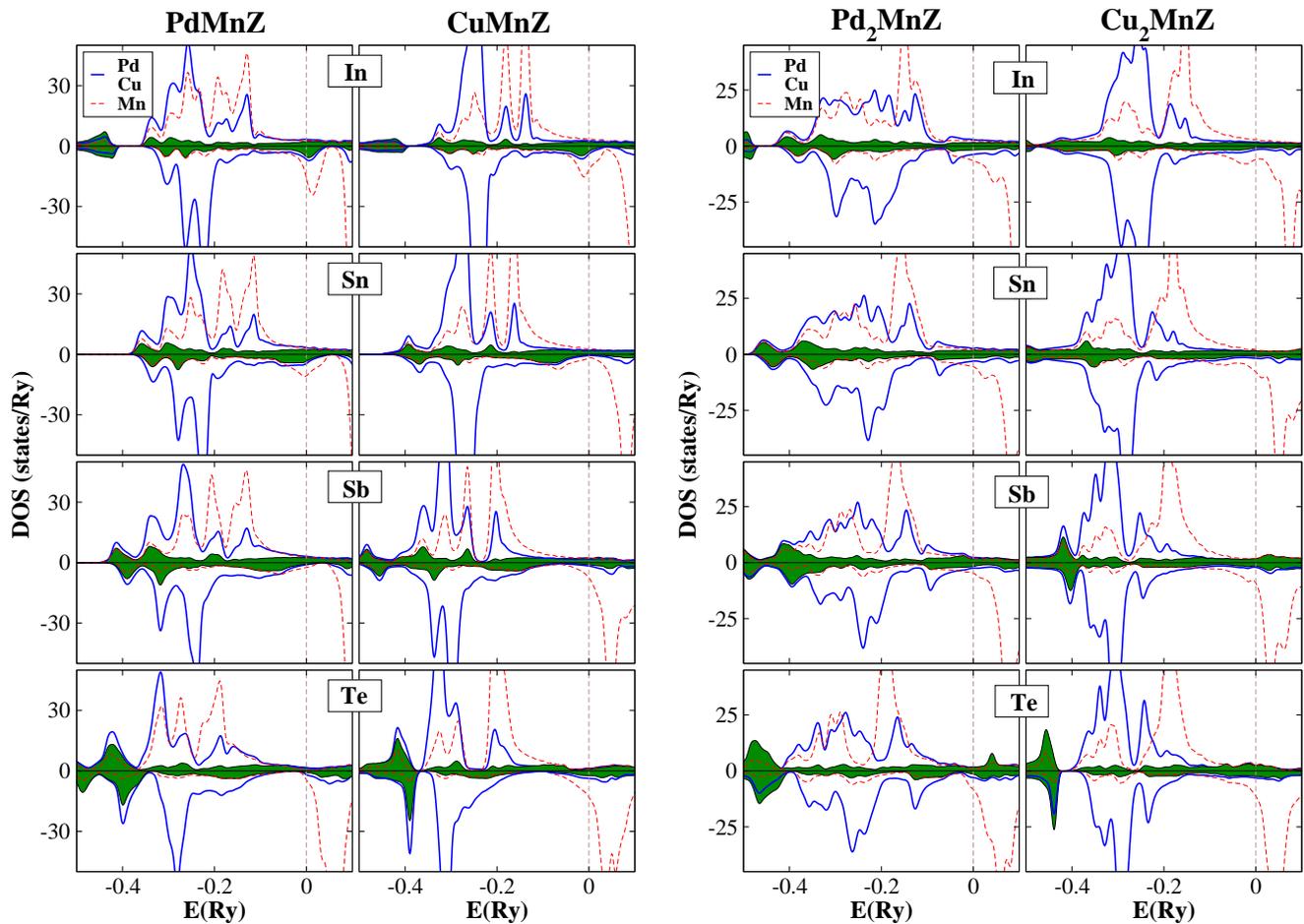

\begin{center}
\includegraphics[scale=0.58]{fig2a.eps}
\hspace{3mm}
\includegraphics[scale=0.58]{fig2b.eps}
\vspace*{-0.4cm}
\end{center}
\caption{(Color online) Left panel: Spin-projected atom-resolved
density of states of PdMnZ and CuMnZ (Z=In, Sn, Sb, Te) for
stoichiometric compositions. The shadow areas show
the  DOS of the Z constituent. The broken vertical lines denote the
Fermi level. Right panel: The same  for  full Heusler  compounds
Pd$_2$MnZ and Cu$_2$MnZ.} \label{fig2}
\end{figure*}

To determine the interatomic exchange interactions we use the
frozen-magnon technique \cite{magnon} and map the results of the
calculation of the total energy of the helical magnetic
configurations\cite{magnon,Heisenberg}
\begin{equation}
\label{spiral} {\bf s}_n=[\cos({\bf qR}_n)\sin{\theta}, \sin({\bf
qR}_n)\sin{\theta}, \cos {\theta}]
\end{equation}
onto a classical Heisenberg Hamiltonian
\begin{equation}
\label{hamiltonian}
 H_{eff}=-  \sum_{i \ne j} J_{ij}
{\bf s}_i{\bf s}_j
\end{equation}
where $J_{ij}$ is an exchange interaction between two Mn sites and
${\bf s}_i$ is the unit vector pointing in the direction of the
magnetic moment at site $i$ , ${\bf R}_n$ are the lattice vectors,
${\bf q}$ is the wave vector of the helix, $\theta$ polar angle
giving the deviation of the moments from the $z$ axis. Within the
Heisenberg model (\ref{hamiltonian}), the energy of frozen-magnon
configurations can be represented in the form
\begin{equation}
\label{eq:e_of_q} E(\theta,{\bf q})=E_0(\theta)-\sin^{2}\theta
J({\bf q})
\end{equation}
where $E_0$ does not depend on {\bf q} and $J({\bf q})$ is the the
Fourier transform of the  parameters of interatomic exchange
interactions:
\begin{equation}
\label{eq:J_q} J({\bf q})=\sum_{\bf R} J_{0{\bf R}}\:\exp(i{\bf
q\cdot R}).
\end{equation}

Calculating $ E(\theta,{\bf q})$ for a regular ${\bf q}$-mesh in
the Brillouin zone of the crystal and performing back Fourier
transformation one gets exchange parameters $J_{0{\bf R}}$ between
pairs of Mn atoms.

The Curie temperature is estimated within the mean-field
approximation (MFA)
\begin{equation}
\label{eq:Tc_MFA} k_BT_C^{MFA}=\frac{2}{3}\sum_{j\ne0}J_{0j}
\end{equation}

\section{Density of States and Magnetic Moments}

The electronic and magnetic  structures of Heusler alloys have
been extensively studied earlier and the reader is referred to
Ref.~\onlinecite{H_2} and the references therein for a detailed
overview. Here we present a brief description of the calculational
results aiming to provide the basis for the discussion of exchange
mechanisms in subsequent sections and to allow the comparison with
previous work.

\subsection{Density of states}

In this section, we discuss the density of states for the
stoichiometric compositions of both families of compounds
[X$_{\textrm{k}}$MnZ ( $\textrm{X}=\textrm{Pd}$, Cu;
$\textrm{k}=1, 2$; $\textrm{Z}=\textrm{In}$, Sn, Sb, Te)]. The
results are presented in Fig.~\ref{fig2}. The DOS for
non-stoichiometric compositions assumes intermediate values. In
agreement with the commonly accepted picture of the magnetism of
the Mn-based Heusler alloys we obtain a strong localization of the
magnetization on the Mn sublattice with a value of the Mn moment
close to 4$\mu_B$. In the following we will show that the value of
the Mn moment is very robust with respect to the variation of the
magnetic structure. Therefore we can conclude that Mn-based
Heusler alloys possess a well-defined atomic Mn moment. The robust
character of the Mn moment results from the large exchange
splitting of the Mn 3\textit{d} states.  Important that the Mn
\textit{3d} states of only one spin projection (spin-up) are
strongly occupied. The main part of the spin-down Mn \textit{3d}
states lies above the Fermi level.

The analysis of the DOS of different compounds reveals a relative
shift of the Fermi level to a higher energy position in the
sequence In-Sn-Sb-Te (Fig.~\ref{fig2}) that results from the
increasing number of valence electrons within this series. Indeed,
a Te atom has three more valence electrons than In, two more
electrons than Sn and one more electron than Sb. The Mn DOS for
semi-Heusler  compounds shows an interesting property that
different compositions give similar features if the total number
of the \textit{sp}-electrons coming from different atoms is the
same. This similarity is  more pronounced near the Fermi level.
For example, the peaks in the Mn spin down states of PdMnSn and
CuMnIn are very similar to each other. As a consequence, the Mn
atomic moments are also similar. This property can be related to
the symmetry of the wave functions in C1$_b$-type crystal
structure of the semi-Heusler alloys.\cite{H_2} In the following
sections we show that the position of the occupied Mn 3\textit{d}
states with respect to the Fermi level, the \textit{sp}-electron
DOS at the Fermi energy, and the structure of the unoccupied part
of the DOS are important for the understanding of the exchange
coupling mechanisms in Heusler alloys.

\subsection{Magnetic moments and local moment behavior}

\begin{figure*}[t]
\begin{center}
\includegraphics[scale=0.585]{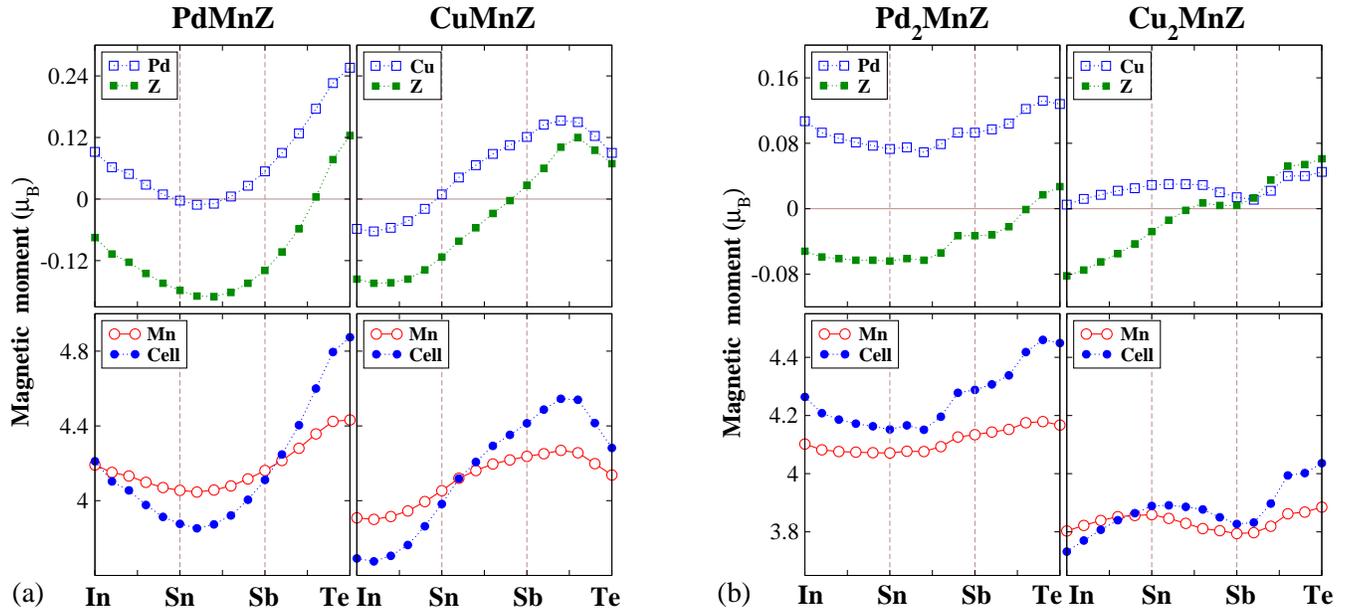}
\vspace*{-0.4cm}
\end{center}
\caption{(Color online) (a) Calculated atom-resolved and total
spin moments (in $\mu_\mathrm{B}$) in PdMnZ and CuMnZ as a
function of the \textit{sp}-electron number of the Z constituent.
(b) The same for full Heusler  alloys Pd$_2$MnZ and Cu$_2$MnZ.}
\label{fig3}
\end{figure*}

Most of the Mn-based systems are believed to possess well-defined
local atomic Mn moments which are usually close to $4 \mu_B$ and
do not change substantially when going from the ordered phase to
the paramagnetic state. The available inelastic neutron scattering
experiments  and magnetization measurements support this point of
view. Early studies of the paramagnetic phase of several
Mn-based compounds have established the absence of spatial
magnetic correlations (spin waves), and have shown the value of
the atomic moment to be in agreement  with the moment obtained
from the static susceptibility measurements. \cite{Heusler_4}
Recently, Plogmann \textrm{et. al.}, gave a detailed study  on the
degree of magnetic moment localization in various Mn-based
full-Heusler alloys using x-ray photoelectron spectroscopy (XPS)
and x-ray emission spectroscopy (XES) techniques.\cite{Heusler_5}
Depending on the atomic number of the Z element, the authors
obtained  from XES experiments an increasing localization of the
Mn-3\textit{d} states that is related to the larger interatomic
distances for heavier \textit{sp} (Z) elements.

On the theoretical side, in 1984 K\"ubler \textit{et al.,} gave a
detailed account of the formation of local moments in various
Mn-based full Heusler alloys using first-principles
calculations.\cite{Heusler_Kubler} Since then many authors have
studied various  semi- and full-Heusler compounds and came to
similar conclusion. An interesting features  arising from the
calculations of Heusler alloys is the resemblance of a number of
physical characteristics such as the position and the width of the
Mn 3\textit{d} peaks, exchange splitting and the value of the Mn
magnetic moment to the corresponding quantities for diluted Mn
impurities in non-magnetic metals.\cite{Local_moment} This
observation gives further evidence for the localized nature of the
magnetism moments in these systems.

\begin{figure*}[t]
\begin{center}
\includegraphics[scale=0.58]{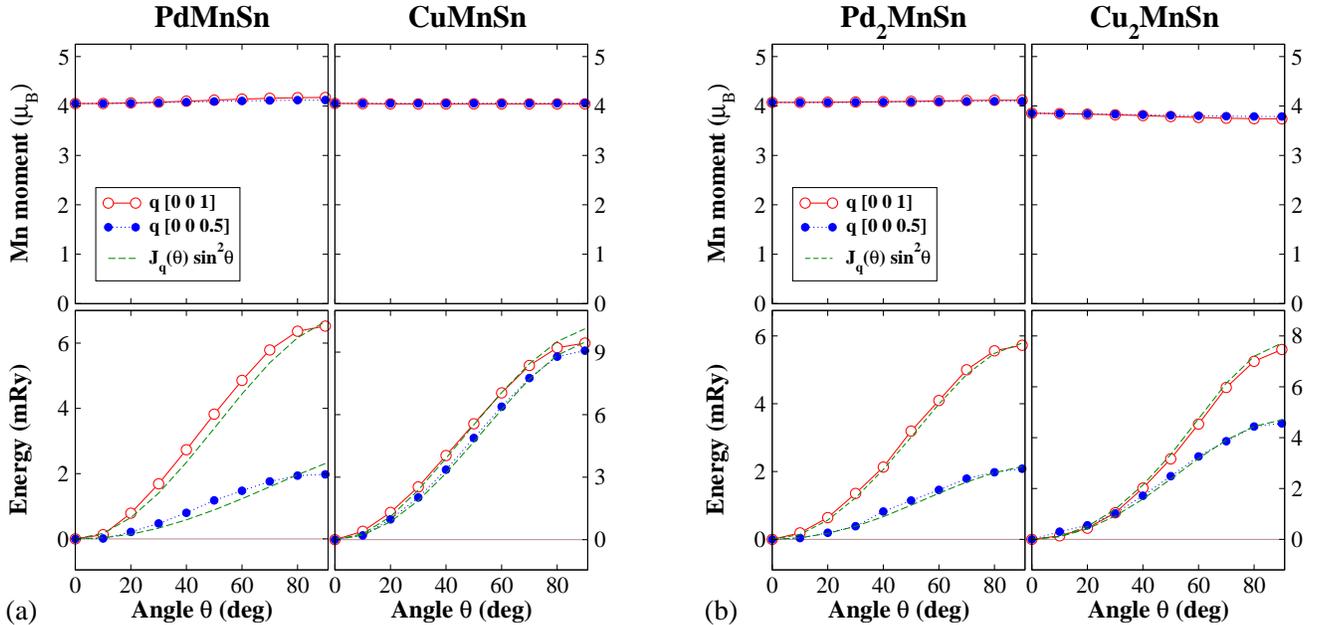}
\vspace*{-0.4cm}
\end{center}
\caption{(Color online) (a) Upper panel: Calculated  Mn spin
moment  in PdMnSn and CuMnSn as  a function of $\theta$ for spin
spiral with $\textbf{q}=(0\:0\:\frac{1}{2}), (0\:0\:1)$ in units
of $2\pi/a$. Lower panel: The corresponding total energies $\Delta
E(\theta, \textbf{q})=E(\theta, \textbf{q})-E(0, 0)$. For
comparison, the results of the force theorem calculations
(broken lines) are presented.
J$_{\textbf{q}}(\theta)$ stand for J$_{\textbf{q}}\times
M_{\theta}^{2}$, $M$ is the magnetic moment. (b) The same for
Pd$_2$MnSn and Cu$_2$MnSn full Heusler compounds.} \label{fig4}
\end{figure*}

In Fig.~\ref{fig3} we  present calculated atom-resolved and total
magnetic moments for both families of  Heusler  alloys PdMnZ,
CuMnZ, Pd$_2$MnZ and Cu$_2$MnZ as a function of the
\textit{sp}-electron number of the Z constituent. As mentioned
above, the magnetic moment is  mostly  confined to the Mn
sublattice. Small moments are induced on the Cu and Pd atoms.
These induced moments are positive (that is parallel to the Mn
moments) in a  broad composition interval. The induced moment of
the Z element is negative. The induced moments as a function of
the Z constituent follow closely the  behavior of the Mn moment.
This correlation is especially well pronounced for semi-Heusler
alloys (Fig.~\ref{fig3}). The comparison of the data for PdMnZ and
CuMnZ shows that for equal total number of valence electrons the
induced moments have similar values. In both semi-Heusler systems,
the variation of the  total magnetic moment with Z constituent is
large. In PdMnZ, it is about 1 $\mu_B$. The half of the variation
comes from the change in the Mn moment. In  full Heusler alloys
the situation is different. Neither a correlation of the
characteristics of the compounds with the same number of the
valance electrons nor a substantial change in magnetic moments
with variation of Z is obtained.

Since an increase of the temperature leads to increasing deviation
of the atomic moments from the direction of the net magnetization
it is important to study the properties of the noncollinear
magnetic configurations. In Fig.~\ref{fig4}, we present the
results of such calculations for four stoichiometric compounds:
PdMnSn, CuMnSn, Pd$_2$MnSn and Cu$_2$MnSn. The results are shown
for the spiral structures with two different wave vectors
$\textbf{q}=(0\:0\:\frac{1}{2})$ and $\textbf{q}=(0\:0\:1)$ in
units of $2\pi/a$ and the polar vector $\theta$ varying in the
interval from 0 to 90$^\circ$ (see Eq. \ref{spiral}).

The upper panels of  Fig.~\ref{fig4} show the $\theta$ dependence
of the Mn moment whereas the bottom panels present the  $\theta$
dependence of the total energy. Note that by the variation of the
wave vector of the spiral and of angle  $\theta$ the magnetic
structure can be continuously transformed from ferromagnetic to
antiferromagnetic. The spirals with $\theta=0$ and arbitrary
\textbf{q}  correspond to the ferromagnetic state whereas the
structure with $\theta=90^\circ$ and $\textbf{q}=(0\:0\:1)$ is
antiferromagnetic.

The analysis of  Fig.~\ref{fig4} shows that for all four compounds
the  magnetic moment of the Mn atom is practically insensitive to
the  magnetic configuration. The relative change of the Mn moment
in transition from ferromagnetic to antiferromagnetic state is
less than 3\%. Therefore the treatment of the Mn moment in these
systems as a local property of the Mn atom is well founded.

The energy of the spiral structures increases monotonously with
increasing $\theta$ and reaches the maximal value for the
antiferromagnetic state. The minimum at $\theta=0$ shows that the
ground state is ferromagnetic. In  Fig.~\ref{fig4}, we compare the
total energy calculated self-consistently with an approximation to
the total energy obtained with the application of so-called
magnetic force theorem \cite{Force_Theorem} The agreement between
two types of calculations is good in the whole $\theta$ interval
that is characteristic for systems with well defined atomic
moments and allows the use of the force theorem for magnetic
configurations deviating strongly from the ground state.

%----------------------------------------------------------------------------------

\section{Exchange Coupling Mechanism} \label{secV}

In this section we suggest an interpretation of the results of the
first-principles calculations on the basis of the Anderson
\textit{s-d} mixing model. The discussion is divided into four
parts. In the first part, the calculated Heisenberg exchange
parameters are presented. In the second part we briefly describe
the Anderson \emph{s-d} mixing model and the exchange mechanisms
resulting from the perturbative treatment of this model. The role
of the conduction-electron spin polarization in exchange coupling
as well as the contribution of the \textit{sp} and non-magnetic
3\textit{d} atoms to the formation of the magnetic state is
discussed in the third part. The last part is devoted to the study
of the influence of non-magnetic 3\textit{d} versus 4\textit{d}
atoms on the exchange coupling.

\subsection{Heisenberg exchange parameters}

\begin{figure*}[t]
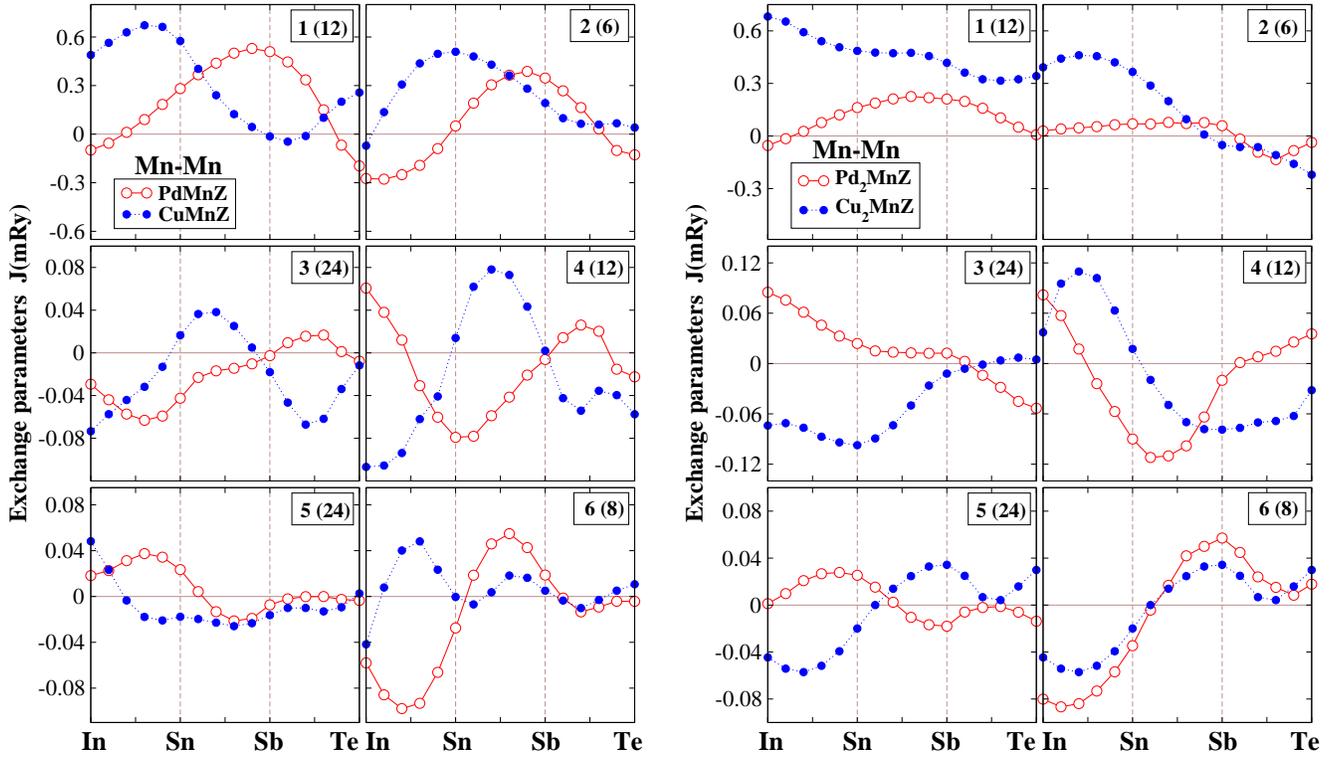

\begin{center}
\includegraphics[scale=0.57]{fig5a.eps}
\hspace{3mm}
\includegraphics[scale=0.57]{fig5b.eps}
\vspace*{-0.3cm}
\end{center}
\caption{(Color online) Left panel: First six nearest neighbor
Mn-Mn exchange parameters in PdMnZ and CuMnZ as a function of the
\textit{sp}-electron number of the Z constituent. Also given are
the number of atoms within corresponding coordination spheres.
Right panel: The same for full-Heusler alloys Pd$_2$MnZ and
Cu$_2$MnZ.} \label{fig5}
\end{figure*}

\begin{figure*}[!ht]
\begin{center}
\includegraphics[scale=0.58]{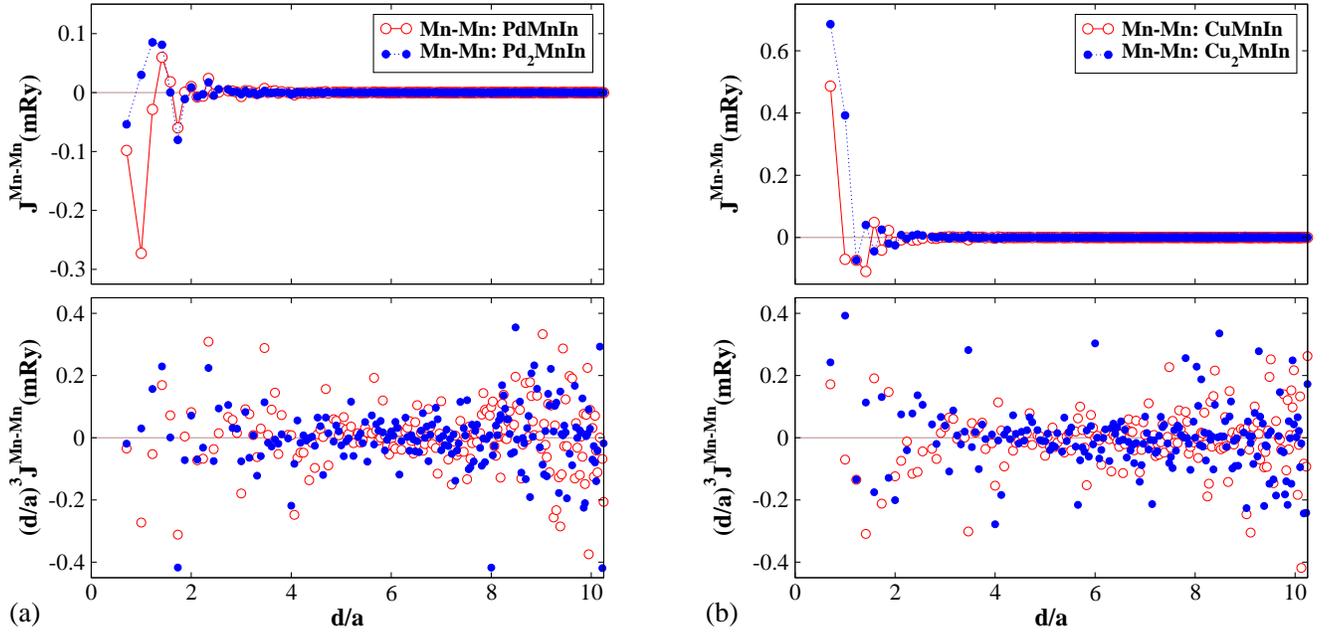}
\vspace*{-0.4cm}
\end{center}
\caption{(Color online) (a) Upper panel: Mn-Mn exchange
interactions  in Pd$_{\textrm{k}}$MnIn  ($\textrm{k}=1,2$) as a
function of distance up to the $10a$. Lower panel: RKKY-type
oscillations in exchange  parameters for corresponding  compounds.
(b) The same for Cu$_{\textrm{k}}$MnIn ($\textrm{k}=1,2$).}
\label{fig6}
\end{figure*}

In Fig.~\ref{fig5}, we present calculated Mn-Mn exchange
parameters for six nearest neighbors as a function of
\textit{sp}-electron concentration. In the upper-right corner of
each panel the number of atoms in the corresponding coordination
sphere is given. The exchange parameters for larger interatomic
distances ($\sim 10\:a$) are shown in Figs.~\ref{fig6} for
selected compounds PdMnIn, CuMnIn, Pd$_2$MnIn and Cu$_2$MnIn. The
absolute value of the exchange parameters decays quickly with
increasing interatomic distance and the main contribution to
$T_{C}$ comes from the interaction between atoms lying closer than
$3a$. No sizable contribution is detected after $5a$. However, the
RKKY-type  oscillations become visible up to very large
interatomic distances when the exchange parameters are multiplied
by $(d/a)^{3}$ (Fig.~\ref{fig6}) where $d$ is the distance between
the interacting Mn atoms and $a$ is the lattice constant).

In agreement with the results  of our previous calculations on
Heusler alloys a strong  dependence of the exchange parameters on
the Z constituent for both families of systems is obtained. As
seen from Fig.~\ref{fig5} all  exchange parameters oscillate
between ferromagnetic and antiferromagnetic values with increasing
\textit{sp}-electron concentration. As discussed below these
oscillations are related to the properties of the electron
structure of the systems.

Considering two nearest neighbor exchange parameters  we notice
that they have ferromagnetic character for a broad composition
range and dominate over the rest of parameters. The remaining
parameters are much weaker. The first two nearest neighbor
exchange interactions are responsible for very high Curie
temperatures  in Cu-based full Heusler alloys as well as in both
classes of semi Heusler alloys.

A distinct feature of the exchange interactions in semi-Heusler
alloys is that the maximum of the exchange interactions for both
PdMnZ and CuMnZ corresponds to the similar numbers of the
\textit{sp}-electrons. This correlates with the conclusions of the
preceding sections where a similar behavior was obtained for the
density of states and the magnetic  moments. The shift of the
maxima for two systems is explained by the fact that Pd has one
\textit{sp}-electron less than Cu. The properties of the exchange
interactions are reflected in the properties of the Curie
temperature (Fig.~\ref{fig7}) where we also obtained a
relative shift of the maxima of the two curves corresponding to
one \textit{sp}-electron. However, no such correlation is obtained
for the full Heusler compounds.

\subsection{Indirect exchange coupling:  RKKY-type exchange and  Superexchange}

In previous calculations on Co$_2$MnSi and Mn$_2$VAl we have shown
that in the systems with several magnetic sublattices the direct
exchange coupling Mn-Co or Mn-V between neighboring  3\textit{d}
atoms can dominate over the indirect Mn-Mn
interactions.\cite{ES_2} For the Mn-based Heusler compounds
considered in this paper the direct coupling does not play
substantial role and can be ignored.\cite{ES_1a} Therefore, in the
following we will consider only the indirect exchange coupling.

The DFT is not based on a model Hamiltonian approach and does not
use a perturbative treatment. Therefore various exchange
mechanisms appear in the results of calculations in a mixed form
that does not allow a straightforward separation of the
contributions of different mechanisms. In this situation, the
model Hamiltonian studies relevant to the problem provide useful
information for a qualitative interpretation of the DFT results.
Among such approaches the Anderson \textit{s-d} mixing model is an
appropriate tool for Mn-based Heusler alloys because of the
localized nature of Mn moments in these systems.

Anderson \textit{s-d} model for a single 3\textit{d} impurity
embedded into a nonmagnetic metallic host is given by the following
Hamiltonian:\cite{Anderson_1}
\begin{eqnarray} \nonumber
H_{s-d}&=&\sum_{\textbf{k}\sigma}\epsilon_{\textbf{k}}a^{\dag}_{\textbf{k}\sigma}
a_{\textbf{k}\sigma}+\sum_{\sigma}\epsilon_{d}n_{d\sigma}\\
\nonumber
&&+\sum_{\textbf{k}\sigma}V_{\textbf{k}d}(a^{\dag}_{\textbf{k}\sigma}
a_{d\sigma}+a_{\textbf{k}\sigma}a^{\dag}_{d\sigma})+Un_{d\uparrow}n_{d\downarrow}
 \end{eqnarray}
where $a^{\dag}_{\textbf{k}\sigma}$ ($a_{\textbf{k}\sigma}$) and
$a^{\dag}_{d\sigma}$ ($a_{d\sigma}$)
create (annihilate) electrons with spin $\sigma$ in the band
states and on the impurity, respectively, $n_{d\sigma}$ is the
number operator for localized electrons of spin $\sigma$, $U$ is
the on-site Coulomb repulsion between two localized electrons
which favors the single occupation of the impurity level.
$V_{\textbf{k}d}$ represents the coupling between the impurity
level and the conduction electrons in the metal. This interaction
causes the mixing  of the band states and impurity level. Such a
mixing gives rise to broadening of the impurity states.
The ratio $U/V$ determines whether a
local magnetic moment in a metallic host can be formed.
The first-principles calculations for 3\textit{d} impurities in
non-magnetic metals gave large values for the magnetic moments
reflecting the strong Coulomb repulsion $U$ between 3\textit{d}
orbitals.\cite{Local_moment} Note that in complex systems like
rare earths and  Heusler alloys one deals with a lattice of magnetic atoms.
In this case, the relevant model is the periodic Anderson \emph{s-d}
model.\cite{PAM_1,PAM_2,PAM_3,PAM_4}

The mixing interaction $V$ induces spin polarization in the
conduction electron sea and spatial propagation of this
polarization gives rise to an effective indirect exchange coupling
between distant magnetic moments. In general, this interaction
contains two distinct processes: electrostatic Coulomb exchange
interaction and \textit{s-d} (or \textit{sp-d}) mixing
interaction. The former induces a net positive spin polarization
while the contribution of the latter is negative and disappears in
strong magnetic limit.\cite{Watson} The indirect exchange coupling
has been described using various theoretical schemes. The Green's
function method and  the perturbation theory are among the most
often used techniques. In 1973 da Silva  and Falicov showed that,
in fourth-order perturbation theory,  the indirect interaction
between two magnetic moments can be separated into two
contributions:\cite{RE_Falicov}
$J_{\textrm{indirect}}=J_{\textrm{RKKY}}+J_{\textrm{S}}$. The
first term is of RKKY type and stems from the intermediate states
which correspond to low energy spin excitations of the Fermi sea,
that is exitations corresponding to electron-hole pair formation
with a spin flip transition. In reciprocal space this contribution
takes the following form.\cite{RE_Falicov,ILEC_1,ILEC_2,ILEC_3}
\begin{eqnarray} \nonumber
J_{\textrm{RKKY}}({\bf q})&=&\sum_{n_1,n_2,k}
\bigg[\frac{|V_{n_{1}k}|^2|V_{n_{2}k'}|^2}{(\varepsilon_{n_{2}k'}-\varepsilon_{d})^2}
\\ \nonumber
& & \times
\frac{\theta(\varepsilon_f-\varepsilon_{n_{1}k})\theta(\varepsilon_{n_{2}k'}-\varepsilon_f)}{\varepsilon_{n_{2}k'}-\varepsilon_{n_{1}k}}
+ c.c. \bigg] \\ \nonumber
\end{eqnarray}
where $\textbf{k}'=\textbf{k}+\textbf{q}+\textbf{G}$ and
\textbf{G} is the reciprocal lattice  vector. This coupling is
largely influenced by the denominator
($\varepsilon_{n_{2}k'}-\varepsilon_{n_{1}k}$). Therefore the topology of
the Fermi surface is an important factor influencing the form of
$J_{RKKY}(\textbf{q})$  with most important contributions coming from the stationary
wave vectors spanning the Fermi surface. These stationary
wave vectors translate into smoothly decaying oscillations in the real space
with a ferromagnetic bias in the "pre-asymptotic" regime.
Also the DOS at the Fermi level, the number of conduction
electrons and their spin polarization play important role in
determining the strength of this interaction.

The second term arises from high-energy virtual charge excitations
in which electrons from local 3\textit{d} states of the magnetic
atom are promoted above the Fermi sea. In the reciprocal space,
the term has the following form that is similar to the form of the
RKKY-type term
\begin{eqnarray} \nonumber
J_{\textrm{S}}({\bf q})&=&-\sum_{n_1,n_2,k}
\bigg[\frac{|V_{n_{1}k}|^2|V_{n_{2}k'}|^2}{(\varepsilon_{n_{2}k'}-\varepsilon_{d})^2}
\\ \nonumber
& & \times
\frac{\theta(\varepsilon_{n_{1}k}-\varepsilon_f)\theta(\varepsilon_{n_{2}k'}-\varepsilon_f)}
{\varepsilon_{n_{1}k}-\varepsilon_d}
+ c.c. \bigg] \\ \nonumber
\end{eqnarray}
In contrast to the RKKY contribution, $J_{\textrm{S}}(\textbf{q})$
does not, however, depend on the DOS at the Fermi level and
the topology of the Fermi surface. Since the sum is taken over
unoccupied states an important role is played by the energy
position of the unoccupied  3\textit{d} states of the  magnetic
atom. The closer the states to the Fermi level the larger
$J_{\textrm{S}}(\textbf{q})$. This interaction is always
antiferromagnetic and  its strength decays exponentially with
distance. A broadening of the 3\textit{d} levels induces weak
oscillations in this coupling. In addition to these parameters,
the position of the occupied 3\textit{d} levels with respect to
the Fermi energy and the strength of the mixing interaction $V$
strongly influence coupling mechanisms.

As shown in Ref.~\onlinecite{ILEC_3} the $\textbf{q} \rightarrow
0$ limit simplifies the above expressions  and is useful for the
qualitative analysis. For $\textbf{q}=0$ the RKKY-type term
becomes.
\begin{equation}
J_{\textrm{RKKY}}(0)=V^4D(\epsilon_{F})/E^{2}_h\nonumber
\end{equation}
where $D(\epsilon_{F})$ is the density of states at the Fermi
level.  $E_h$ is the energy required to promote an electron from
occupied 3\textit{d} level to the Fermi level. Parameter $E_h$ is
not well defined in Heusler alloys because of the broadening of
the Mn 3\textit{d} levels into the energy bands crossing the Fermi
level. In all alloys studied the occupied Mn 3\textit{d}  peaks
lie below $-0.1$ Ry where the energy is counted off the Fermi level
(Fig.~\ref{fig2}).
On the other hand, the  $\textbf{q}=0$ limit of the superexchange term cannot be 
expressed in terms of the density of states and has more complex form
\begin{equation}
J_{\textrm{S}}(0)=V^4\sum_{nk}[\epsilon_F-\epsilon_{nk}-E_h]^{-3}\nonumber
\end{equation}

In addition, we would like to comment on the following two points.
First, although the perturbative derivation of above expressions
is based on the assumption of two magnetic impurities embedded
into metallic host, the generalization to the periodic lattices is
straightforward. As shown by  G. Malmstr\"{o}m \textit{et. al.},
and Price  the final results differ by a phase
factor.\cite{Heusler_Malmstrom_1,Heusler_Price}. Second, there are
two different limits in the description of the exchange mechanisms
within Anderson \textit{s-d} model. In weak magnetic limit where
the coupling is dominated by the \textit{s-d} mixing of the local
and conduction electron states, both mechanisms mentioned above
coexist and their relative contributions are determined by the
details of the electronic structure of the system. In the opposite
limit (strong magnetic limit) in which the coupling is primarily
due to the electrostatic Coulomb exchange interaction, the second
term is not present and the first term is reduced to the
conventional RKKY interaction.\cite{ILEC_3} The systems considered
in this paper are characterized by strong \textit{s-d}
hybridization and therefore we expect the presence of the
contributions of both exchange mechanisms.

\subsection{Conduction electron spin polarization}

\begin{figure*}[t]
\begin{center}
\includegraphics[scale=0.58]{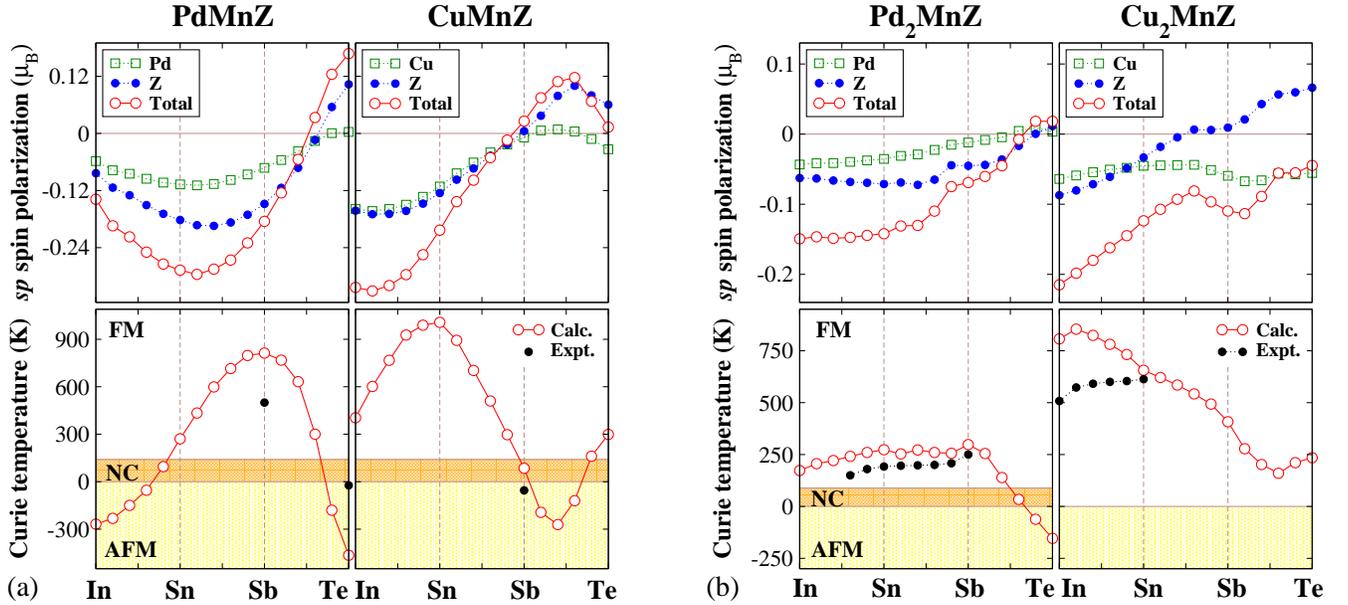}
\vspace*{-0.3cm}
\end{center}
\caption{(Color online)  (a) Upper panel: The \textit{sp}-electron
spin polarization (in $\mu_\mathrm{B}$) of the X=Pd,Cu and Z
constituents. The total polarization is given as the sum of X and
Z spin polarizations. Lower  panel: Mean-field estimation of the
Curie temperature in PdMnZ and CuMnZ. For comparison available
experimental $T_{C}$ values taken from Refs.\onlinecite{Heusler_1}
and \onlinecite{Heusler_2} are presented. FM, NC and AFM stand for
ferromagnetic, non-collinear, and antiferromagnetic ordering,
respectively . (b) The same for the full-Heusler  alloys Pd$_2$MnZ
and Cu$_2$MnZ. } \label{fig7}
\end{figure*}

Now we turn to the interpretation of the DFT results in terms of
the two exchange mechanisms discussed in the previous section. To
estimate the relative contribution of the ferromagnetic RKKY-type
exchange we present in Fig.~\ref{fig7} the calculated conduction
electron spin polarization in both families of Heusler alloys as a
function of the electron number of the Z constituent. The analysis
of the calculational data allows us to draw a number of important
conclusions. First, in both families of alloys the direction of
the induced spin polarization is opposite to the direction of the
Mn moment in a broad interval of compositions. This feature
reveals the primary role of the \textit{sp-d} mixing in the
exchange coupling and justifies the use of the Anderson
\textit{s-d} model for the description of the magnetism in these
systems.

Another remarkable feature is a very clear correlation between the
spin polarization and the mean-field Curie temperature (or
exchange parameters) in a large part of the phase diagram
(Fig.~\ref{fig7}). Indeed, the compounds with very large spin
polarization are characterized by the value of the Curie
temperature that is also very high. Interestingly, for the zero
polarization the Curie temperature also vanishes or assumes very
small values reflecting the dominating character of the
ferromagnetic RKKY-type exchange mechanism in establishing
magnetic order. However, at some regions superexchange mechanism
becomes important. This can be seen in Fig.~\ref{fig7} where for
PdMnIn$_{1-x}$Sn$_{x}$ ($x< 0.8$) system we obtain an
antiferromagnetic order in spite of very large spin polarization.
Further insight into the nature of the coupling can be gained from
Fig.~\ref{fig2} where  we present atom and spin resolved density
of states of both families of compounds for stoichiometric
compositions. As pointed out above, the superexchange mechanism is
sensitive to the DOS above the Fermi energy. As seen in
Fig.~\ref{fig2} the Mn 3\textit{d} states provide the main
contribution to the DOS in this region. For PdMnIn the DOS peak
above the Fermi level assumes the largest value and, as a result
the antiferromagnetic superexchange dominates over the
ferromagnetic RKKY-type exchange giving rise to AFM order. In
transition from In to Sn the peak gradually decreases and
therefore the superexchange becomes less important. Around
$\textrm{Z}=\textrm{Sb}$ it almost disappears leading to FM order.
However, another large peak approaches the Fermi level for
$\textrm{Z}=\textrm{Te}$ that turns the systems from a ferromagnet
into an antiferromagnet with a non-collinear ordering in the
intermediate region due to the competition of two mechanisms. The
situation is very similar in the case of CuMnZ where we also
obtain a rich magnetic behavior.

On the other hand, for the full-Heusler alloys the magnetic phase
diagram is rather simple compared to the semi-Heusler systems. In
Cu$_2$MnZ, the ground state is ferromagnetic for all Z. The
mean-field T$_C$ qualitatively follows the behavior of conduction
electron spin polarization, assuming largest value for
$\textrm{Z}=\textrm{In}$ and gradually decreasing in the
In-Sn-Sb-Te sequence. Around Sb the conduction electron spin
polarization increases that in principle should lead to the
increase of the Curie temperature according to the discussion of
the previous section. However, in this region the T$_\textmd{C}$
further decreases  revealing substantial contribution of the
antiferromagnetic superexchange. This is reflected in position of
the Mn 3\textit{d} states that is very close to the Fermi level
(Fig.~\ref{fig2}).

In the Pd-based full Heusler alloys Pd$_2$MnZ the spin polarization is
rather small, and as a result the Curie temperature is low. The
spin polarization in In-Sn interval is independent of the Z
constituent, while in the rest of the phase diagram it gradually
decreases  and becomes zero around Te. As for the T$_{\textrm{C}}$
we  obtain a similar Z-independent behavior in a large interval of
compositions (from In to Sb). However, from Sb to Te the
T$_\textrm{C}$ sharply decreases due to dominating contribution of
superexchange. The calculated T$_\textrm{C}$ values are in good
agreement with the experiments for the Pd-based full Heusler
alloys, while they are overestimated in PdMnSb and Cu-based full
Heusler compounds. The following explanation of this property can
be suggested. In the Pd-based full Heusler alloys the more distant
exchange parameters contribute with a substantial weight to the
formation of the Curie temperature (Fig.~\ref{fig5}) whereas in
PdMnSb and Cu-based full Heusler compounds only the first and
second nearest neighbor parameters determine the T$_\textrm{C}$.
In the latter case the mean-field approximation is less exact and
overestimates the Curie temperature. For such systems, the
random-phase approximation (RPA) is expected to provide a better
description of the T$_\textrm{C}$.

The results of the first principles calculations and the
qualitative analysis on the basis of the Anderson \emph{s-d} model
allow us to formulate two conditions for high Curie temperature.
(i) The conduction electron spin polarization should be maximal.
(ii) The DOS above Fermi level should be minimal.  Indeed, as seen
from Figs.~\ref{fig2} and \ref{fig7}, the compounds CuMnSn, PdMnSb
and Cu$_2$MnIn that satisfy both conditions posses very high Curie
temperatures. The knowledge of the conditions for high Curie
temperature is an important help in the fabrication of the
materials with desired properties. From the point of view of the
second condition the half-metallic ferromagnets have an advantage
since the gap in the spin-down channel decreases the number of the
states just above the Fermi level. Indeed, a large number of
first-principles calculations showed that these materials have
very high Curie temperatures.
\cite{Kubler2003,Mark,DMS,Sakuma,Sanyal,MnSi,Sato}

As mentioned in preceding part the direction of the conduction
electron spin polarization with respect to local moment and its
amplitude are important in classifying ferromagnets. From the
obtained results the full-Heusler alloys can be characterized as
weak ferromagnets since spin polarization is negative for all Z
constituents, i.e., \textit{sp-d} mixing is dominating. As for the
semi-Heusler alloys the situation is similar to the full-Heusler
alloys except some regions in the magnetic phase diagram where
spin polarization changes sign: Sb-Te interval in CuMnSb and the
region around Te in PdMnZ. A positive spin polarization does not
directly mean that these alloys are strongly ferromagnetic. They
are, however, close to this limit and the Mn-Mn coupling is
primarily due to the electrostatic Coulomb exchange interaction.
Indeed, as seen from Figs.~\ref{fig3} and \ref{fig7} the alloys
with large positive spin polarization  have large Mn magnetic
moments. In PdMnTe both quantities assume the largest values:
m$_{\textrm{Mn}}=4.4 \mu_B$ and m$_{\textrm{sp}}=0.18 \mu_B$. The
dependences of the Mn moment and of the spin polarization on the Z
constituent are correlated (Figs.~\ref{fig3} and \ref{fig7}).

\begin{figure*}[t]
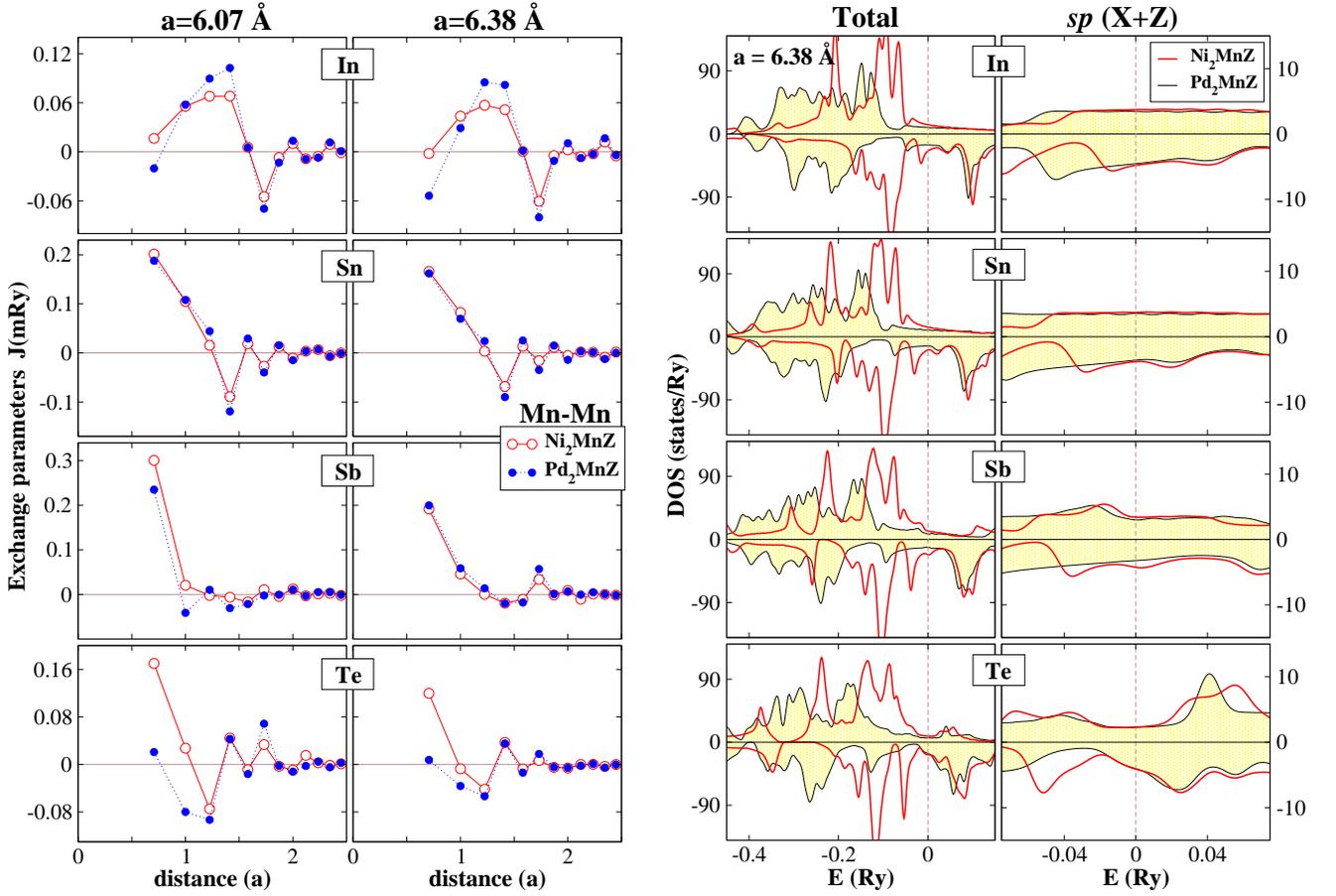

\begin{center}
\includegraphics[scale=0.56]{fig8a.eps}
\hspace{3mm}
\includegraphics[scale=0.56]{fig8b.eps}
\vspace*{-0.5cm}
\end{center}
\caption{(Color online) Left panel: The parameters of the Mn-Mn
exchange interactions in Ni$_2$MnZ and Pd$_2$MnZ ($\textrm{Z=In}$,
Sn, Sb, Te) for two different lattice constants. Right panel: Spin
projected total DOS (left hand side)  and \textit{sp}-electron DOS
of X and Z atoms (right hand side) for Ni$_2$MnZ and Pd$_2$MnZ
($\textrm{Z=In}$, Sn, Sb, Te) for the lattice constants of $6.38
\AA$.} \label{fig8}
\end{figure*}

Because of the importance of the conduction-electron spin
polarization a direct comparison with experimental measurement of
this quantity is desirable. An important information on
conduction-electron spin polarization is supplied by the
measurements of the hyperfine fields at non-magnetic sites (X,Z).
The strength of the transferred hyperfine fields correlates with
the amplitude of the \emph{s} conduction electron spin
polarization. Indeed, the measurements by Campbell and Khoi
\emph{et. al.}, showed that maximal \textit{s}-electron spin
polarization is found in the systems with high Curie temperatures
such as Cu$_2$MnAl and Cu$_2$MnIn.\cite{Hyperfine_1,Hyperfine_2}
However, the hyperfine fields are sensitive to only the \emph{s}
electron spin polarization and do not give information about the
polarization of the \emph{p} electrons that are usually dominating
in Heusler alloys. For probing the total conduction electron spin
polarization the magnetic Compton scattering profiles are proved
to be a useful tool. Using this method, Zukowski \emph{et. al.}
obtained recently a large \emph{sp}-electron spin polarization in
Cu$_2$MnAl which is antiferromagnetically coupled to the Mn
moments.\cite{polarization_1} A similar result is obtained  by Deb
\emph{et. al.}, for Ni$_2$MnSn.\cite{polarization_2} Our
calculations are in agreement with both experiments.

\subsection{3d versus 4d electrons}

\begin{table*}
\begin{center}
\caption{Magnetic moments, \textit{sp}-electron polarization (in
$\mu_B$) and Curie temperatures in  Ni$_2$MnZ  and Pd$_2$MnZ
($\textrm{Z}=\textrm{In}$, Sn, Sb, Te) for two different lattice
parameters. Negative values of T$_\textrm{C}$  mean that ground
state is antiferromagnetic.}
\begin{ruledtabular}
\begin{tabular}{lcccccc|cccccc}
&\multicolumn{5}{c}{$a=6.07$ \AA}&\multicolumn{5}{c}{$a=6.38$
\AA}\\\hline
 Compound  & X & Mn  &  Z & \textit{sp} & Cell & T$_{\textrm{C}}^{\textmd{MFA}}$(K) & X & Mn  &  Z & \textit{sp} &
 Cell & T$_{\textrm{C}}^{\textmd{MFA}}$(K)
\\ \hline
     Ni$_2$MnIn    & 0.27  & 3.72 & -0.07 & -0.16  & 4.19 & 254 & 0.27  & 3.99 & -0.07 & -0.17  & 4.45 & 165 \\
     Pd$_2$MnIn    & 0.11  & 3.81 & -0.05 & -0.14  & 3.98 & 279 & 0.17  & 4.10 & -0.05 & -0.15  & 4.26 & 173 \\
\hline
     Ni$_2$MnSn    & 0.21  & 3.74 & -0.06 & -0.13  & 4.10 & 321 & 0.21  & 4.00 & -0.06 & -0.14  & 4.35 & 251 \\
     Pd$_2$MnSn    & 0.09  & 3.79 & -0.06 & -0.13  & 3.91 & 368 & 0.07  & 4.07 & -0.06 & -0.14  & 4.15 & 263 \\
\hline
     Ni$_2$MnSb    & 0.15  & 3.76 & -0.03 & -0.09  & 4.04 & 309 & 0.22  & 4.04 & -0.02 &  -0.06 & 4.46 & 234 \\
     Pd$_2$MnSb    & 0.08  & 3.82 & -0.03 & -0.08  & 3.95 & 193 & 0.09  & 4.13 & -0.03 & -0.07  & 4.29 & 296 \\
\hline
     Ni$_2$MnTe    & 0.22  & 3.83 &  0.03 & 0.01  & 4.31  & 120 & 0.21  & 4.04 &  0.02 & -0.03  & 4.48 & 46 \\
     Pd$_2$MnTe    & 0.14  & 3.89 &  0.04 & 0.04  & 4.21  & -172 & 0.13  & 4.17 &  0.03 &  0.02 & 4.45 & -154 \\
\end{tabular}
\end{ruledtabular}
\label{table2}
\end{center}
\end{table*}

In majority of the Heusler alloys with chemical formula X$_2$YZ
and XYZ the Y site is occupied by the Mn atom, while for the X
site there is much more freedom: here can be any element from
3\textit{d}, 4\textit{d}, or 5\textit{d} late transition metals
(i.e. Fe, Ru, Co, Rh, Ni, Pd, Pt, Cu, Ag and Au). The magnetic
moment of the X atom can for many alloys be neglected. The
exceptions are some of the 3\textit{d} atoms, e.g., Fe or Co.
Since the delocalization of the \textit{d} states increases with
transition from  3\textit{d} elements to  4\textit{d} and further
to 5\textit{d} elements one can expect a substantial dependence of
the magnetic properties of the Heusler alloys for the occupation
of the X site with atoms from different \textit{d} series. The
purpose of this section is to study this dependence by means of
comparison of two systems: Ni$_2$MnZ and Pd$_2$MnZ
($\textrm{Z}=\textrm{In}$, Sn, Sb, Te).

We choose two different lattice parameters for each system:
6.07$\AA$ and 6.38$\AA$. The first corresponds to the lattice
constant of Ni-based alloys while the second is the characteristic
for the Pd-based systems. The both systems are experimentally well
studied. In Fig.~\ref{fig8} we present spin projected total DOS
and \textit{sp}-electron DOS of the X and Z atoms for the lattice
constant of 6.38$\AA$. Note that Ni (3\textit{d}) is isoelectronic
to Pd (4\textit{d}). The calculated magnetic moments are given in
Table~\ref{table2}. The comparison of the results obtained for the
same lattice constant reveals similarity of some features: the
value of the Mn magnetic moment, \textit{sp}-electron spin
polarization and DOS of the \textit{sp} states of X and Z atoms
around the Fermi level (see Fig.~\ref{fig8} and
Table~\ref{table2}). On the other hand, well below the Fermi
energy  the DOS of the Ni-based and Pd-based systems differ
strongly. The peaks of the DOS of the Ni-based compounds are
higher and narrower than in the Pd-based systems. This is mainly
the result of a more delocalized character of the Pd 4\textit{d}
orbitals compared to the Ni 3\textit{d} orbitals. As we will show,
despite the strong difference of the occupied part of the DOS the
magnetic properties of two systems are rather close that gives
additional support to the conclusion that the states lying close
to the Fermi level play the most important role in the formation
of magnetic properties.

In Fig.~\ref{fig8} (left panel) we present the parameters of the
Mn-Mn exchange interactions for Ni$_2$MnZ and Pd$_2$MnZ
($\textrm{Z=In}$, Sn, Sb, Te) for two different values of the
lattice constant. The Curie temperatures estimated within
mean-field approximation are given in Table~\ref{table2}. For both
systems the patterns of exchange parameters for the equal lattice
constants are rather similar. The parameters are very close to
each other for $\textrm{Z}=\textrm{Sn}$ and Sb. For
$\textrm{Z}=\textrm{Te}$, the difference is stronger. This
difference is more pronounced for near neighbors. For example the
difference in the first two exchange parameters makes Pd$_2$MnTe
antiferromagnetic in contrast to the ferromagnetic  Ni$_2$MnTe.

\begin{figure}[t]
\begin{center}
\includegraphics[scale=0.59]{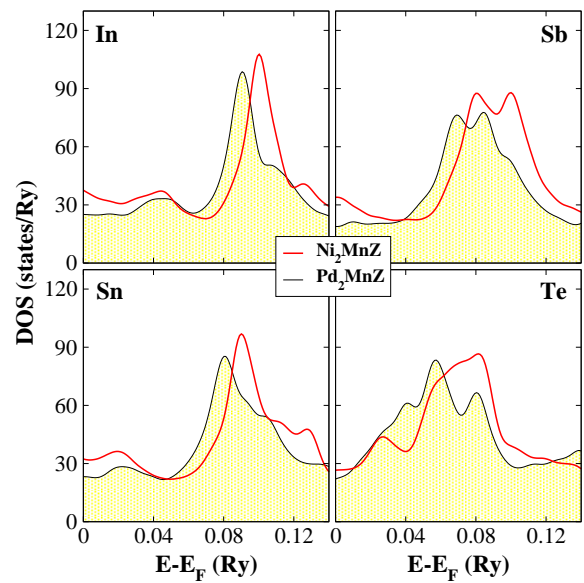}
\vspace*{-0.4cm}
\end{center}
\caption{(Color online) The total DOS of Ni$_2$MnZ and Pd$_2$MnZ
(Z$=$In, Sn, Sb, Te) above Fermi level.} \label{fig9}
\end{figure}

The analysis in terms of the competition between two types of
exchange interactions suggested in the previous section is helpful
also in this case. The long range behavior of the exchange
parameters, specifically the RKKY-type oscillations can be related
to \textit{sp}-electron spin polarization and total
\textit{sp}-electron DOS at the Fermi level. As seen from
Table~\ref{table2} and Fig.~\ref{fig8} these quantities are very
close to each other in both systems that explains the similarity
in the long range behavior of the exchange interactions. However,
the short range behavior depends very much on the contribution of
the antiferromagnetic superexchange. To reveal the importance of
this mechanism we present in Fig.~\ref{fig9} the total DOS
(spin-up plus spin-down) above the Fermi energy for both systems
and for the lattice constant of 6.38$\AA$. As seen in
Fig.~\ref{fig9}, the DOS above the Fermi level is not identical
for $\textrm{X}=\textrm{Ni}$ and $\textrm{X}=\textrm{Pd}$.
Therefore the deviations in exchange parameters at least within
few coordination spheres can be expected. A very strong similarity
in the patterns of the exchange parameters for Sn and Sb alloys
compared to the In and Te systems cannot be explained on the basis
of the differences in the DOS and might be accidental resulting
from the cancellation of the contributions of different states to
the superexchange.

Next, we comment  on the volume dependence of the magnetic
properties. As seen in Table~\ref{table2} and  Fig.~\ref{fig8},
the compression of the volume (the reduction of the lattice
parameter) leads to the decrease of the Mn magnetic moments and an
increase of the absolute value of the leading exchange parameters.
The reduction of the magnetic moments is an expected result which
is a consequence of the increased interatomic hybridization and
broadening of the electron bands. The volume dependence of the
exchange interactions is less straightforward. In general this
dependence is non-monotonous and in concrete physical situations
one can find both an increase and a decrease of the exchange
interactions with volume contraction. A detailed discussion of
these aspects can be found in Ref.~\onlinecite{ES_1b} where the
electronic structure, exchange interactions and Curie temperature
of the full-Heusler alloy Ni$_2$MnSn are studied as a function of
pressure. It was shown that in low pressure region the
T$_\textrm{C}$ behavior is in qualitative correlation with
empirical interaction curve of Kanomata \textit{et. al.} which
describes the dependence of the Curie temperature of the Mn-based
Heusler alloys on the Mn-Mn distance.\cite{Pressure_1} In
agreement with experiment we have found that at ambient pressure
T$_\textrm{C}$ increases with increasing pressure that is
$dT_{\textrm{C}}/dP>0$. \cite{Pressure_1,Pressure_2,Pressure_3}
The pressure dependence has a maximum at 3.6 \AA. Indeed, as seen
in Table~\ref{table2} in agreement with experiments we obtain the
same behavior in the T$_\textrm{C}$ (T$_\textrm{N}$ for
Pd$_2$MnTe) for all compounds except Pd$_2$MnSb in which the Curie
temperature decreases with reduction of Mn-Mn distance. To
summarize, the magnetism of the Heusler alloys containing
different \textit{d} (3\textit{d} or 4\textit{d}) electrons
appeared to be qualitatively similar whereas quantitative
differences result from the stronger delocalization of the
4\textit{d} electrons and larger lattice parameters of the
compounds having 4\textit{d} elements.

\section{summary and conclusions}

We performed a systematic first-principles study to reveal the
exchange mechanisms in various Mn-based semi- and full-Heusler
alloys. The calculation of the exchange parameters is based on the
frozen-magnon approach and  the Curie temperature is estimated
within the mean-field approximation. Due to large separation of
the Mn atoms and the local moment nature of magnetism in these
systems the exchange coupling is indirect and is mediated by the
conduction electrons. The results obtained are interpreted using
\textit{s-d} mixing model of Anderson. To understand the
dependence of physical characteristics on the valence electron
number we go beyond the stoichiometric compositions employing
virtual crystal approximation. The influence of non-magnetic
3\textit{d} versus 4\textit{d} electrons on exchange coupling is
discussed.

We show that magnetism in these systems strongly depends on the
number of conduction electrons, their spin polarization and the
position of the unoccupied Mn 3\textit{d} states with respect to
Fermi level. Various magnetic phases are obtained depending on the
combination of these characteristics. The magnetic phase diagram
is determined at $\textrm{T}=0$. We find that in the case of a
large conduction electron spin polarization and the unoccupied Mn
3\textit{d} states lying far above the Fermi level, an RKKY-type
ferromagnetic interaction is dominating. On the other hand, the
antiferromagnetic superexchange becomes important in the presence
of large peaks of the unoccupied Mn 3\textit{d} states lying close
to the Fermi energy. The resulting magnetic behavior depends on
the competition of these two exchange mechanisms. The obtained
results are in good correlation with the conclusions made on the
basis of the Anderson \textit{s-d} model and with available
experimental data. These findings suggest that a targeted
influence on the corresponding physical quantities can provide a
useful tool for the fabrication of materials with desired physical
properties.

\begin{acknowledgments}
The financial support of Bundesministerium f\"ur Bildung und
Forschung is acknowledged.
\end{acknowledgments}

\end{document}